\input{hyperlatex}
\documentstyle[preprint,eqsecnum,aps]{revtex}

\input epsf

\begin{document}

\preprint{\vbox{
\hbox{UCSD/PTH 96--19}
}}
\title{Relating CKM Parametrizations and Unitarity Triangles}
\author{Richard F. Lebed\footnote{rlebed@ucsd.edu}}
\address{Department of Physics, University of California at
San Diego, La Jolla, CA 92093}
\date{July, 1996}
\maketitle
\begin{abstract}

	Concise variable transformations between the four angles of
the CKM matrix in the standard representation advocated by the
Particle Data Group and the angles of the unitarity triangles are
derived.  The behavior of these transformations in various limits is
explored.  The straightforward extension of this calculation to other
representations and more generations is indicated.
\end{abstract}



\narrowtext

\section{Standard Parametrization vs.\ Unitarity Angles}

	The most popular model for parametrizing quark flavor-changing
currents and CP violation is that of mixing between quark mass and
weak interaction eigenstates, as represented by the unitary
Cabibbo--Kobayashi--Maskawa (CKM) matrix\cite{CKM}.  Although the
unitary $N \times N$ matrix for $N$ quark generations possesses
$(N-1)^2$ observable real parameters, these parameters may be (and
have been) chosen in countless different ways.  Even if we adopt the
usual prescription of $N(N-1)/2$ Euler rotation angles and
$(N-1)(N-2)/2$ phases in generation space, we are still faced with the
choice of which axes to use for our rotations and in what order to
perform them; this choice leads to no less than 36 distinct but
equivalent parametrizations for three generations\cite{J1}.  The
particular form of the CKM matrix advocated by the Particle Data
Group\cite{PDG}, as originally proposed by Chau and Keung\cite{CK}, is
just one of these, and is written
\begin{equation} \label{std}
V_{\rm CKM} = \left(
\begin{array}{ccc}
c_{12} c_{13} & s_{12} c_{13} & s_{13} e^{ -i \delta_{\scriptstyle
13}} \\ -s_{12} c_{23} - c_{12} s_{23} s_{13} e^{i
\delta_{\scriptstyle 13}} & c_{12} c_{23} - s_{12} s_{23} s_{13} e^{i
\delta_{\scriptstyle 13}} & s_{23} c_{13} \\ s_{12} s_{23} - c_{12}
c_{23} s_{13} e^{i \delta_{\scriptstyle 13}} & -c_{12} s_{23} - s_{12}
c_{23} s_{13} e^{i \delta_{\scriptstyle 13}} & c_{23} c_{13}
\end{array}
\right) ,
\end{equation}
where $c_{ij} \equiv \cos \theta_{ij}$ and $s_{ij} \equiv \sin
\theta_{ij}$, the subscripts indicate the plane of rotation, and
the Euler angles $\theta_{12}$, $\theta_{23}$, and $\theta_{13}$ are
all chosen to lie in the first quadrant by a redefinition of
(unobservable) quark field phases.  The phase angle $\delta_{13}$ may
not be similarly restricted:
\begin{equation}
0 \leq \theta_{ij} \leq \frac{\pi}{2}, \hspace{2em} 0 \leq \delta_{13}
< 2\pi .
\end{equation}

	Alternately, the CKM matrix may be described in terms of
parameters invariant under choice of convention or phase
redefinitions.  The moduli of the elements $|V_{\alpha i}|$ fall into
this category but are not always the most convenient variables in
experimental measurements\cite{J2}.  For example, the short-distance
contribution to $B\bar B$ mixing, ubiquitous in neutral $B$ decays, is
proportional to $|V_{td} V^*_{tb}|^2$.  Unitarity information may be
more easily recovered by noting that $V V^\dagger = 1$ is equivalent
to the orthogonality of columns or rows in $V$:
\begin{eqnarray} \label{orth}
\sum_{\alpha = u,c,t, \ldots} V_{\alpha i} V^*_{\alpha j} & = &
\delta_{ij}, \nonumber \\ \sum_{i = d,s,b, \ldots} V_{\alpha i}
V^*_{\beta i} & = & \delta_{\alpha \beta} .
\end{eqnarray}
For two distinct columns or rows, the right hand side is zero, and so
the condition may be depicted geometrically in complex space as
describing a closed polygon with a one side corresponding to each
quark generation.  The conditions from~(\ref{orth}) with 1 on the
right hand side set the scale of the polygons.  For three generations
one obtains triangles, which are special since knowledge of the angles
is sufficient to determine their shapes uniquely; thus we concentrate
on the three-generation case in this work.  Equation~(\ref{orth})
implies that there are 6 independent triangles, called the unitarity
triangles\cite{BJJarl}, which are pictured in Fig.~1 and labeled by
the pair of rows or columns whose orthogonality is represented.

	Our chief interest in the unitarity triangles is that their
angles are convention-independent.  One sees this by noting that for
two complex numbers $z_1$ and $z_2$, the (oriented exterior) angle
between them is $\arg(z_1^* z_2)$, where the argument function assumes
its principal value, $-\pi < \arg(z) \leq \pi$, for all complex $z$.
Thus angles in this case have the form
\begin{equation} \label{angle}
\omega^{ij}_{\alpha \beta} \equiv \arg (V_{\alpha i} V^*_{\alpha j}
V_{\beta j} V^*_{\beta i}) .
\end{equation}
Each quark index in this expression appears in both a $V$ and a $V^*$,
so that any phase redefinition cancels in the product.  Geometrically,
the redefinition of a quark phase simply rotates an entire unitarity
triangle by a constant angle.

	It is also known that the 6 unitarity triangles all have the
same area, given by the Jarlskog parameter $J$\cite{J3}.  This follows
first because $3 \times 3$ unitary matrices enjoy the property that a
a particular pattern of elements is invariant up to a sign:
\begin{equation} \label{jdef}
{\rm Im} (V_{\alpha i} V^*_{\alpha j} V_{\beta j} V^*_{\beta i}) = J
\sum_{\gamma, k} \epsilon_{\alpha \beta \gamma} \, \epsilon_{ijk} ,
\end{equation}
which defines $J$ for any choice of $\alpha \neq \beta$, $i \neq j$,
and second because the area of a triangle with sides $z_1$ and $z_2$
is $|{\rm Im} (z_1^* z_2)| /2 = |z_1| \cdot |z_2| \cdot \left| \sin
\left[ \arg(z_1^* z_2) \right] \right|/2$.  It follows that the area
of each unitarity triangle is given by $|J|/2$.  $J$ is the unique
convention-independent CP violation parameter of the CKM matrix, in
that all measurable CP-violating quantities turn out to be
proportional to $J$, so that vanishing area in any of the unitarity
triangles indicates the vanishing of CP violation in the CKM matrix.

{\vspace{4ex}\multiply\baselineskip
by 3 \divide\baselineskip by 4
\vbox{\vbox{\hfil\epsfbox{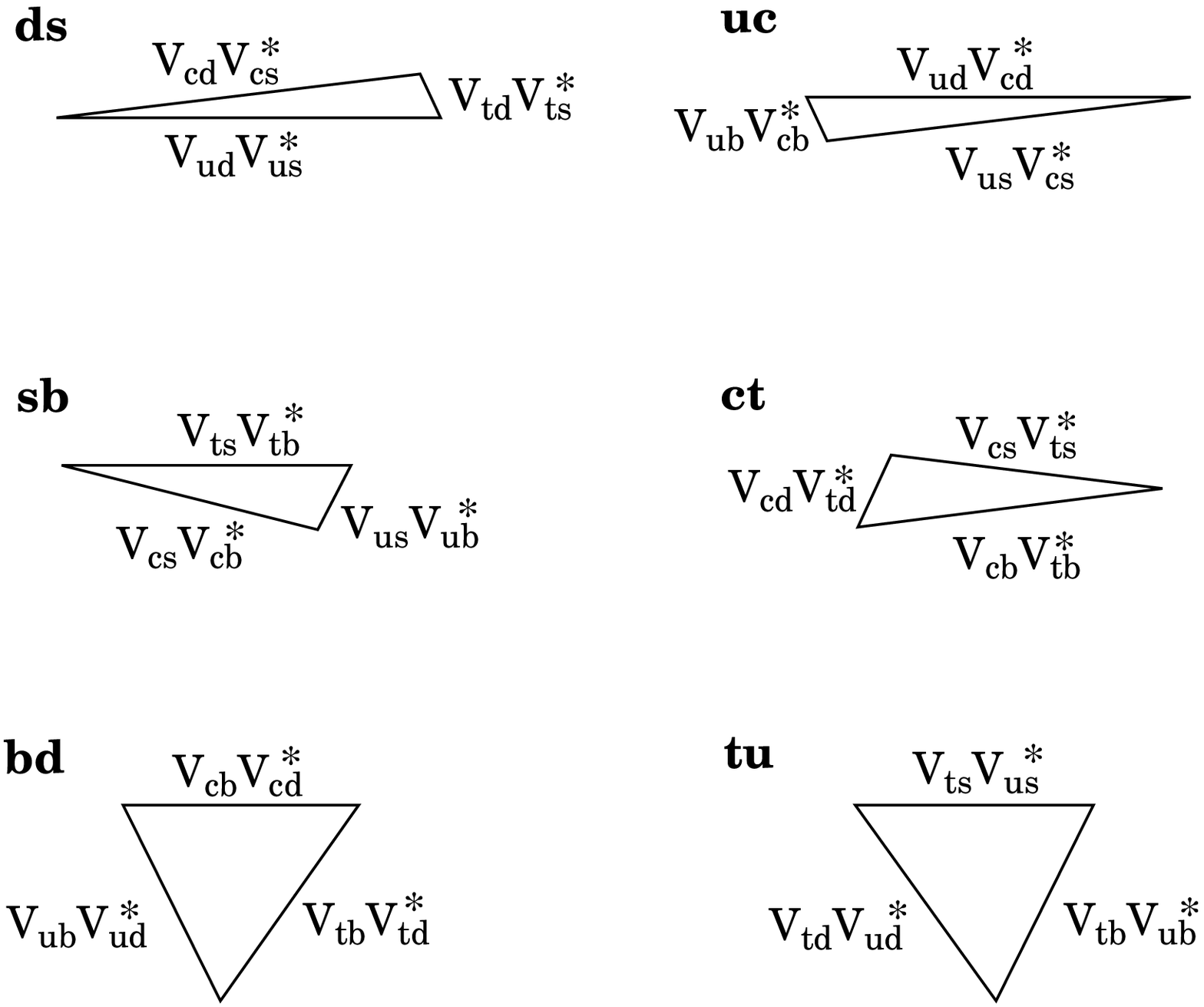}\hfill}%
{
\noindent%
%
{FIG.\ 1. }{\sl The unitarity triangles for three quark generations,
as presented in Ref.~\cite{AKL}.  The triangles are labeled by the
pair of rows or columns whose orthogonality is represented.  The
angles as numbered are: $1 \equiv \alpha$; $2 \equiv \beta$; $3 \equiv
\pi - (\alpha + \beta)$ (conventionally called $\gamma$ if closure of
the triangle is not assumed); $4 \equiv \beta + \epsilon -
\epsilon^\prime$; $5 \equiv \pi - (\alpha + \beta + \epsilon -
\epsilon^\prime)$; $ 6 \equiv \epsilon$; $7 \equiv \alpha + \beta -
\epsilon^\prime$; $8 \equiv \pi - (\beta + \epsilon)$; $9 \equiv
\epsilon^\prime$.  In all cases, the arrows on the complex vectors are
oriented counterclockwise, indicating the experimental positivity of
$J$. \medskip}}
}\multiply\baselineskip by 4 \divide\baselineskip by 3}%

	Phenomenologically, much is already known about the unitarity
triangles.  As schematically indicated in Fig.~1, four of them are
nearly flat because they possess one side much shorter than the other
two.  Of the remaining two, the {\bf bd} triangle is of greatest
current interest to experimentalists: Its sides, which are expected to
be of comparable length, represent the least well-known elements of
the CKM matrix and will be accessible in several $B$ factories
currently under development.  There is already a large
literature dedicated to methods of measuring the sides of
the {\bf bd} triangle and extracting its angles, denoted $\alpha$,
$\beta$, and $\gamma$; see Ref.~\cite{rev} for a recent review.

\section{A Complete Set of Unitarity Angles}

	All that has been said up to this point is already several
years old.  What has been appreciated only recently is that the angles
of the unitarity triangles enjoy a number of elegant properties, and
may be used\cite{AKL} to reconstruct the full CKM matrix except for
the sign of $J$, which has been determined by experiment to be
positive\cite{sgnJ}.  The success of this approach follows from the
observation that the 6 triangles have 18 distinct sides in all, but
only 9 distinct angles.  One sees this by noticing that
$\omega^{ij}_{\alpha \beta}$ appears in both the {\bf ij} and
{\boldmath $\alpha \beta$\unboldmath} triangles, halving the potential
number; furthermore, applying the condition that the angles of any
given triangle add to $\pi$ shows that these 9 angles may be written
as sums and differences of only 4 independent angles.  Since this is
also the number of independent parameters in the three-generation CKM
matrix, the angles may be used as a basis for all
convention-independent quantities.

	To be specific, we follow~\cite{AKL} in defining the interior
angles:
\begin{eqnarray} \label{aklang}
\alpha & \equiv & \pi - | \omega^{bd}_{tu} | \nonumber \\
\beta & \equiv & \pi - | \omega^{bd}_{ct} | \nonumber \\
\epsilon & \equiv & \pi - | \omega^{sb}_{ct} | \nonumber \\
\epsilon^\prime & \equiv & \pi - | \omega^{ds}_{uc} | .
\end{eqnarray}
These definitions are used to label the angles in Fig.~1.  In
particular, $\alpha$ and $\beta$ (not to be confused with the indices
of $\omega^{ij}_{\alpha \beta}$) are the same angles traditionally
used in the literature for the {\bf bd} triangle, so these angles form
a natural and very useful basis for describing invariants of the CKM
matrix.  The form of these expressions is chosen so that each of
$\alpha$, $\beta$, $\epsilon$, $\epsilon^\prime$ lies in the range
$(0,\pi)$.  Note that the quantities $\omega^{ij}_{\alpha \beta}$ as
defined in Eq.~(\ref{angle}) appear different from, but are formally
identical to, those in Ref.~\cite{AKL}, since $\arg(z_1/z_2^*) =
\arg(z_1 z_2)$.

	Observe that, although the parameters $\omega^{ij}_{\alpha
\beta}$ do indeed contain the sum total of the unitarity information,
some information is lost in the definitions (\ref{aklang}), since
magnitudes of $\omega^{ij}_{\alpha \beta}$ are taken.  In fact, all
that is lost is the orientation of the angles, namely, whether they
are constructed in the clockwise or counterclockwise direction.  Since
the angles must still form closed triangles, this formulation merely
surrenders one's ability to distinguish between a particular triangle
and its mirror image.  This freedom corresponds to the sign of $J$, as
indicated by Eqs.~(\ref{angle}) and (\ref{jdef}).

	Reference~\cite{AKL} shows how to reconstruct all measurable
phases and moduli of the CKM matrix given only the angles of the
unitarity triangles.  On the other hand, all of the parametrizations
of the CKM matrix perform the same function, but in terms of
quantities interpreted in a convention-dependent way.  In the
remainder of this paper, we derive a complete set of concise relations
between the two parametrizations for the particular standard form of
Eq.~(\ref{std}).  Then we check that the transformations obey the
appropriate limits for vanishing CP violation in the CKM matrix and
consider the transformations in a phenomenologically useful
limit. Finally, we indicate the straightforward generalization to
other parametrization choices and more quark generations.

\section{The Transformations}

	First observe that the following relation holds for all
complex $z$:
\begin{equation} \label{cotdef}
\cot (\pi - | \arg (z) | ) = -\frac{{\rm Re}(z)}{|{\rm Im} (z)|} .
\end{equation}
Since the argument function is related to the arctangent function, it
is clear that one will obtain the cleanest expressions for the
unitarity angles in terms of their tangents or cotangents; why
cotangents are superior for our purposes will become evident.  In
fact, simply inserting the elements of (\ref{std}) into
Eqs.~(\ref{angle}) and (\ref{aklang}) obtains the desired
transformations in one direction for all angles of the unitarity
triangles.  We use the following compact notation: $s$, $c$, $t$, and
$c \hspace{-1ex} /$ respectively represent sine, cosine, tangent, and
cotangent, while the indices $x$, $y$, $z$, and $\delta$ respectively
represent $\theta_{12}$, $\theta_{23}$, $\theta_{13}$, and
$\delta_{13}$.  A number in the index indicates a multiple angle so
that, for example, $s_{2y} \equiv \sin 2\theta_{23}$.  Cotangents of
the 9 angles (or their supplements) appearing in the unitarity
triangles are:
\begin{eqnarray} \label{ltor}
\cot \alpha & = & \frac{+c \hspace{-1ex} /_x c \hspace{-1ex} /_y s_z -
c_\delta}{|s_\delta|} , \nonumber \\
\cot (\alpha + \beta) & = & \frac{-c \hspace{-1ex} /_x t_y s_z -
c_\delta}{|s_\delta|} , \nonumber \\
\cot (\alpha + \beta - \epsilon^\prime) & = & \frac{+t_x t_y s_z -
c_\delta}{|s_\delta|} , \nonumber \\
\cot (\alpha + \beta + \epsilon - \epsilon^\prime) & = & \frac{-t_x c
\hspace{-1ex} /_y s_z - c_\delta}{|s_\delta|} , \nonumber \\
\cot \beta & = & \frac{(s^2_x - c^2_x s^2_z) s_{2y} - s_{2x} c_{2y}
s_z c_\delta}{s_{2x} s_z |s_\delta|} , \nonumber \\
\cot \epsilon & = & \frac{(c^2_x - s^2_x s^2_z) s_{2y} + s_{2x} c_{2y}
s_z c_\delta}{s_{2x} s_z |s_\delta|} , \nonumber \\
\cot \epsilon^\prime & = & \frac{(c^2_y - s^2_y s^2_z) s_{2x} + c_{2x}
s_{2y} s_z c_\delta}{s_{2y} s_z |s_\delta|} , \nonumber \\
\cot ( \beta + \epsilon - \epsilon^\prime ) & = & \frac{(s^2_y - c^2_y
s^2_z) s_{2x} - c_{2x} s_{2y} s_z c_\delta}{s_{2y} s_z |s_\delta|} ,
\nonumber \\
\cot( \beta + \epsilon) & = & \frac{\left\{ s^2_{2x} \left[
\frac{1}{4} s^2_{2y} ( 1+ s^2_z )^2 - s^2_z \right] - \frac{1}{4}
s_{4x} s_{4y} s_z ( 1 + s^2_z ) c_\delta - s^2_{2y} s^2_z ( 1 -
s^2_{2x} c^2_\delta ) \right\}}{s_{2x} s_{2y} s_z c^2_z |s_\delta|} .
\nonumber \\ & &
\end{eqnarray}

	Several comments are in order.  First, it is obvious that the
first four of these expressions are quite simple, the next four are of
intermediate complexity, and the last is quite complicated.  The
origin of this distinction becomes clear with a glance at (\ref{std}):
The elements with complicated forms in the lower-left $2 \times 2$
submatrix (each the sum of a real and a complex number) respectively
appear 1, 2, or 4 times in using (\ref{angle}) to compute the
corresponding first four, middle four, and final expressions in
Eq.~(\ref{ltor}).  This feature is repeated in any parametrization of
the CKM matrix using Euler angles and phases.  Since the angles
appearing in the first four expressions in (\ref{ltor}) are
independent, and the cotangent function is one-to-one on $(0,\pi)$,
these four simple expressions contain all the information of the CKM
matrix except the sign of $J$.  The chosen CKM parametrization picks
out particular combinations of the basis angles $\alpha$, $\beta$,
$\epsilon$, and $\epsilon^\prime$ in which the transformation
equations are simple.  It is now clear that cotangents are chosen over
tangents so that adding the quantities in (\ref{ltor}) in order to
invert them is simpler.

	Second, as for the sign of $J$, note that the dependence of
each expression in (\ref{ltor}) on the CP-violating phase
$\delta_{13}$ occurs only through the functional forms $\cos
\delta_{13}$ or $|\sin \delta_{13}|$ and so is insensitive to the
variable change $\delta_{13} \to (2\pi - \delta_{13})$; this is
explicitly how the parametrization of (\ref{std}) is sensitive to the
sign of $J$ but the angles of Ref.~\cite{AKL} are not.

	Finally, important limiting cases are evident from these
transformations.  As is well known, CP violation does not occur in the
CKM matrix if any of the following conditions hold:
\begin{equation} \label{cond}
\theta_{ij} = 0, \, \frac{\pi}{2} \hspace{0.5em} \mbox{\rm for any of}
\hspace{0.5em} ij = 12, \, 23, \, 13 ; \hspace{2em} \delta_{13} = 0,
\, \pi .
\end{equation}
In such cases our transformations (\ref{ltor}) must satisfy the
property that the unitarity angles collapse to zero area.  To see
this, note that the denominator of each expression in (\ref{ltor}), as
seen from Eqs.~(\ref{angle}), (\ref{jdef}), and (\ref{cotdef}), is
just $|J|$.  In the standard parametrization~(\ref{std}),
\begin{equation}
J =  s_{12} c_{12} s_{23} c_{23} s_{13} c_{13}^2
s_{\delta_{\scriptstyle 13}} ,
\end{equation}
so that $J$ is seen to vanish when any of the conditions in
(\ref{cond}) is satisfied.  If one could ignore the numerators in
(\ref{cotdef}), each expression in (\ref{ltor}) would become singular
under the conditions (\ref{cond}), making each unitarity angle 0 or
$\pi$ so that the unitarity triangles would collapse.  However, in
some cases the numerator factors cancel factors in the denominator,
and so some of the angles continue to assume finite values even when
certain conditions in (\ref{cond}) are satisfied.  For example, for
$\theta_{23} \to \pi/2$, $\cot \alpha \to -\cos \delta_{13} / | \sin
\delta_{13} | \neq \infty$.  In such cases, however, the unitarity
triangles may still be seen to collapse.  For, in the given example,
consider the {\bf bd} triangle in Fig.~1.  In the same limit, $\cot
\beta \to +\cos \delta_{13} / | \sin \delta_{13} |$ and $\cot (\alpha
+ \beta) \to -\infty$, so that $\alpha$ and $\beta$ add to $\pi$.
Thus the sides $|V_{ub} V^*_{ud}|$ and $|V_{cb} V^*_{cd}|$ are
parallel, requiring $|V_{tb} V^*_{td}|$ to have zero length for the
triangle to close, and the triangle collapses.  All limiting cases
from Eq.~(\ref{cond}) lead to trivial unitarity triangles using
similar observations.

	The angles $\alpha$, $\beta$, $\epsilon$, and
$\epsilon^\prime$ as defined in (\ref{aklang}) are bounded between 0
and $\pi$.  However, not all values for all of the angles are
simultaneously allowed.  In order for the unitarity triangles to
close, their allowed ranges must be correlated.  If one chooses them
in the order $\alpha$, $\beta$, $\epsilon$, $\epsilon^\prime$, one
requires
\begin{eqnarray} \label{range}
& & 0 < \alpha < \pi , \nonumber \\
& & 0 < \beta < \pi - \alpha , \nonumber \\
& & 0 < \epsilon < \pi - \beta , \nonumber \\
& & \max ( 0, \alpha + \beta + \epsilon  - \pi ) < \epsilon^\prime <
\beta + \min ( \alpha , \epsilon ) .
\end{eqnarray}

	We now derive the inverse transformations to Eq.~(\ref{ltor}).
As pointed out above, four combinations of the angles are particularly
convenient to work with for this purpose, and for convenience we use
the following notation for them:
\begin{eqnarray}
A & \equiv & \cot \alpha , \nonumber \\
B & \equiv & \cot(\alpha + \beta) , \nonumber \\
C & \equiv & \cot(\alpha + \beta - \epsilon^\prime) , \nonumber \\
D & \equiv & \cot(\alpha + \beta + \epsilon - \epsilon^\prime) .
\end{eqnarray}
{}From (\ref{ltor}) one sees that differences of $A$, $B$, $C$, and
$D$ eliminate the $\cos \delta_{13}$ factor, and quotients eliminate
the $| \sin \delta_{13}|$.  In particular,
\begin{eqnarray}
A-B & = & c \hspace{-1ex} /_x s_z (c \hspace{-1ex} /_y + t_y) /
|s_\delta| , \nonumber \\
A-D & = & c \hspace{-1ex} /_y s_z (c \hspace{-1ex} /_x + t_x) /
|s_\delta| , \nonumber \\
C-D & = & t_x s_z (c \hspace{-1ex} /_y + t_y) / |s_\delta| , \nonumber
\\
C-B & = & t_y s_z (c \hspace{-1ex} /_x + t_x) / |s_\delta| .
\end{eqnarray}
The quantities given here are manifestly nonnegative, as may be
checked using the range constraints (\ref{range}) and the monotonic
decrease of cotangent over $(0, \pi)$.  One other combination is
particularly simple:
\begin{equation}
BD-AC = s_z c_\delta (c \hspace{-1ex} /_x + t_x)(c \hspace{-1ex} /_y +
t_y) / |s_\delta|^2 .
\end{equation}
From here it is trivial to obtain expressions for the parameters of
the standard form (\ref{std}):
\begin{eqnarray} \label{rtol}
\cot \theta_{12} \equiv c \hspace{-1ex} /_x & = &
\sqrt{\frac{A-B}{C-D}} , \nonumber \\
\cot \theta_{23} \equiv c \hspace{-1ex} /_y & = &
\sqrt{\frac{A-D}{C-B}} , \nonumber \\
\sin \theta_{13} \equiv s_z & = &
\sqrt{\frac{(A-B)(A-D)(C-B)(C-D)}{(A-B+C-D)^2+(BD-AC)^2}} , \nonumber
\\
\cos \delta_{13} \equiv c_\delta & = & \frac{BD-AC}{\sqrt
{(A-B+C-D)^2+(BD-AC)^2}} .
\end{eqnarray}
It is permissible to use only the positive branches of square roots
since the angles $\theta_{12}$, $\theta_{23}$, $\theta_{13}$ in
(\ref{std}) are chosen to lie in the first quadrant, and cotangent and
sine are one-to-one on $(0,\pi/2)$.  Note that only $\cos \delta_{13}$
can be determined, reflecting the discrete ambiguity between
$\delta_{13}$ and $(2 \pi - \delta_{13})$.  One may be troubled by the
fact that parameters specific to a particular representation of the
CKM matrix are written here in terms of its invariants, but this
indicates only that the interpretation of a given parameter as a phase
or rotation about particular axes, not its value, is
convention-dependent.  The situation is analogous to Lorentz
invariance: The norm of the momentum four-vector of a free particle is
most easily computed in the rest frame, where the zero component of
the vector has the interpretation of rest mass $m$, but $m$ is also
numerically the norm of the vector in any frame.

	One can also see explicitly from the transformations
(\ref{rtol}) how the CP-vanishing cases of (\ref{cond}) are recovered
when the unitarity triangles collapse.  If the angles $\alpha$,
$\beta$, $\epsilon$, or $\epsilon^\prime$ approach 0 or $\pi$, then
the cotangents either become singular, as for $A$ when $\alpha \to 0,
\, \pi$, or two of them become degenerate, such as $A$ and $B$ when
$\beta \to 0, \, \pi$.  In such cases, the expressions in (\ref{rtol})
are seen to assume the values of Eq.~(\ref{cond}).  For completeness,
we present the Jarlskog parameter in these variables, which is seen to
satisfy the same degeneracy constraints:
\begin{equation}
|J| = \frac{(A-B)(A-D)(C-B)(C-D) \left[ (A-B)(1+CD) + (C-D)(1+AD)
\right]}{\left[ (A-B+C-D)^2 + (BD-AC)^2 \right]^2} .
\end{equation}

	Finally, we consider the transformations in the
phenomenologically interesting case of $\epsilon^\prime \ll \epsilon
\ll 1$.  It is straightforward to show from Eqs.~(\ref{rtol}) that
\begin{eqnarray}
\theta_{12} & = & \sqrt{ \cot \beta - \cot ( \alpha + \beta )} \,
\epsilon^{1/2} + O ( \epsilon^{3/2} ) , \nonumber \\
\theta_{23} & = & \sqrt{ \cot \beta - \cot ( \alpha + \beta )} \,
\epsilon^{\prime \, 1/2} + O ( \epsilon^{\prime \, 1/2} \epsilon ) ,
\nonumber \\
\theta_{13} & = & \frac{(\epsilon^\prime \epsilon)^{1/2}}{\sin
(\alpha + \beta)} + O ( \epsilon^{\prime \, 1/2} \epsilon^{3/2} ) ,
\nonumber \\
\delta_{13} & = & \pi - {\rm sgn} (J) ( \alpha + \beta -
\epsilon^\prime ) + O( \epsilon^\prime \epsilon ).
\end{eqnarray}
Note the particular vanishing behaviors of the angles as $\epsilon$,
$\epsilon^\prime \to 0$.  Also note that in this limit $\delta_{13} =
\pi - \alpha - \beta = \gamma$, using the experimentally determined
positive sign of $J$.

\section{Other Parametrizations}

	{}From the construction detailed above, it should be clear
that an analogous program can be carried out for any parametrization
using Euler angles and phases, leading to concise transformation
expressions.  The key point is that in such parametrizations elements
more complicated than a real number times a phase, which complicate
the calculation in Eq.~(\ref{angle}), are relegated to a $2 \times 2$
minor submatrix.  The unique element of the CKM matrix that does not
share a row or column index with this $2 \times 2$ minor is thus
distinguished; in the case of the standard form (\ref{std}), this
element is $V_{ub}$.  The angles of the unitarity triangles for which
the analogous expressions to (\ref{ltor}) are simple are exactly those
adjacent to a side containing the distinguished element of $V$, as is
clear for the case we have considered, from Fig.~1.  As another
example, consider the original CKM parametrization of Kobayashi and
Maskawa\cite{CKM}:
\begin{equation}
V_{\rm KM} = \left(
\begin{array}{ccc}
c_1 & -s_1 c_3 & -s_1 s_3 \\
s_1 c_2 & c_1 c_2 c_3 - s_2 s_3 e^{i\delta} & c_1 c_2 s_3 + s_2 c_3
e^{i\delta} \\
s_1 s_2 & c_1 s_2 c_3 + c_2 s_3 e^{i\delta} & c_1 s_2 s_3 - c_2 c_3
e^{i\delta}
\end{array}
\right) .
\end{equation}
Here the distinguished element is $V_{ud}$, and the simplest
expressions to use will be the cotangents of $\alpha$, $(\alpha +
\beta)$, $\epsilon^\prime$, and $(\beta + \epsilon -
\epsilon^\prime)$.

	A similar treatment for the Wolfenstein
parametrization\cite{Wolf}, defined by
\begin{equation}
V_{\rm W} = \left(
\begin{array}{ccc}
1 -\frac{1}{2} \lambda^2 & \lambda & A \lambda^3 (\rho - i\eta) \\
-\lambda & 1 - \frac{1}{2} \lambda^2 & A \lambda^2 \\
A \lambda^3 ( 1 - \rho - i \eta ) & -A \lambda^2 & 1
\end{array}
\right) ,
\end{equation}
is not immediately possible, since the matrix is only unitary to
corrections of order $\lambda^4$.  Only once an extension rendering it
fully unitary, of which there are many possible choices, is agreed
upon can the full conversion between $A$, $\lambda$, $\rho$, and
$\eta$ and the unitarity angles be carried out.

	Finally, we show how any four independent moduli $|V_{\alpha
i}|$ form an equivalent set to the four independent unitarity angles
or standard CKM angles modulo the sign of $J$.  Showing they are
equivalent to the unitarity angles requires using the normalization of
rows and columns of the CKM matrix to compute the other five moduli,
and using the full set to compute the lengths of the sides (and hence
the angles) of the unitarity triangles.  Starting instead with the
standard form (\ref{std}), one computes the remaining five moduli as
before and then extracts $\theta_{13}$, $\theta_{23}$, and
$\theta_{12}$ respectively from $|V_{ub}|$, $|V_{cb}|/|V_{tb}|$, and
$|V_{us}|/|V_{ud}|$.  $\delta_{13}$ may be extracted from any of the
four moduli in the lower-left $2 \times 2$ submatrix, but because a
number and its complex conjugate have the same norm, this process is
insensitive to the transformation $\delta_{13} \to (2 \pi -
\delta_{13})$, or equivalently the sign of $J$.  Similar remarks apply
to any particular representation.

	If it turns out that there are more than three generations of
quarks, the angles of the unitarity polygons are no longer sufficient
to determine the entire structure of the CKM matrix.  In four
generations, for example, a square and a rectangle have the same
angles but are not similar figures.  Nor are moduli alone enough, as
one sees from comparing rhombi and squares.  It is clear that one then
requires both unitarity angles and moduli.  Nevertheless,
generalizations of Euler forms like Eq.~(\ref{std}) to more than three
generations must continue to have ``distinguished'' elements for which
expressions relating convention-dependent CKM angles to
convention-independent angles of unitarity polygons remains usefully
succinct.

\vskip3em
{\it Acknowledgments}
\hfil\break
This work was supported by the Department of Energy under contract
DOE-FG03-90ER40546.


\begin{thebibliography}{99}

\bibitem{CKM} N. Cabibbo, Phys.\ Rev.\ Lett.\ {\bf 10} (1963) 531;
M. Kobayashi and T. Maskawa, Prog.\ Theor.\ Phys.\ {\bf 49} (1973)
552.

\bibitem{J1} C. Jarlskog, in {\it CP Violation}, edited by C. Jarlskog
(World Scientific, Singapore, 1989), p. 3.

\bibitem{PDG} Particle Data Group (L. Montanet {\it et al.}), Phys.\
Rev.\ D {\bf 50} (1994) 1173.

\bibitem{CK} L.-L. Chau and W.-Y. Keung, Phys.\ Rev.\ Lett.\ {\bf 53}
(1984) 1802.

\bibitem{J2} Nevertheless, it is possible to reconstruct all four
invariant quantities of the CKM matrix (but not the sign of the
Jarlskog parameter) using only four independent moduli $|V_{\alpha
i}|$: see C. Jarlskog and R. Stora, Phys.\ Lett.\ B {\bf 208} (1988)
268.

\bibitem{BJJarl} The idea of picturing unitarity conditions in this
way appears to have originated independently with J. D. Bjorken and
C. Jarlskog (see paper in~\cite{J2}).

\bibitem{J3} C. Jarlskog, Phys.\ Rev.\ Lett.\ {\bf 55} (1985) 1039.

\bibitem{rev} A. Ali, Report No.\ DESY 96-106 [hep-ph/9606324]
(unpublished).

\bibitem{AKL} R. Aleksan, B. Kayser, and D. London, Phys.\ Rev.\
Lett.\ {\bf 73} (1994) 18.

\bibitem{sgnJ} In fact, the sign of $J$ appears to be already
experimentally determined as +1.  In the Wolfenstein parametrization
$J \approx A^2 \lambda^6 \eta$, and $\eta$ as determined through $K
\bar K$ mixing is positive, assuming the bag parameter ${\hat B}_K$
from short-distance calculations is also positive (see, for example,
Ref.~\cite{rev}).  $\eta$ could have been negative, for example, if
the neutral $K$ eigenstate mass difference $\Delta m$ had been
negative.

\bibitem{Wolf} L. Wolfenstein, Phys.\ Rev.\ Lett.\ {\bf 51} (1983)
1945.

\end{thebibliography}
\end{document}